\documentclass[aps,prl,superscriptaddress,floatfix]{revtex4}
\usepackage{amsfonts}
\usepackage{epsfig}
\usepackage{amsmath,amssymb}
\usepackage{times}
\usepackage{xcolor}
\usepackage{multirow}
\usepackage{tabularx}
\usepackage{textcomp}

\begin{document}
%\begin{CJK*}{GBK}{kai}
%\preprint{APS/123-QED}

\title{Effective information spreading based on local information in correlated networks}

\author{Lei Gao}
\affiliation{Web Sciences Center, University of Electronic
Science and Technology of China, Chengdu 610054, China}
\affiliation{Big data research center, University of Electronic Science and Technology of China, Chengdu 610054, China}

\author{Wei Wang}
\affiliation{Web Sciences Center, University of Electronic
Science and Technology of China, Chengdu 610054, China}
\affiliation{Big data research center, University of Electronic Science and Technology of China, Chengdu 610054, China}

\author{Liming Pan}
\affiliation{Web Sciences Center, University of Electronic
Science and Technology of China, Chengdu 610054, China}
\affiliation{Big data research center, University of Electronic Science and Technology of China, Chengdu 610054, China}

\author{Ming Tang\footnote{Correspondence to: tangminghan007@gmail.com}}
\affiliation{Web Sciences Center, University of Electronic
Science and Technology of China, Chengdu 610054, China}
\affiliation{Big data research center, University of Electronic Science and Technology of China, Chengdu 610054, China}

\author{Hai-Feng Zhang\footnote{Correspondence to: haifengzhang1978@gmail.com}}
\affiliation{School of Mathematical Science, Anhui University, Hefei 230601, China}

\date{\today}

\begin{abstract}
Using network-based information to facilitate information spreading is an essential task for spreading dynamics in complex networks, which will benefit the promotion of technical innovations, healthy behaviors, new products, etc. Focusing on degree correlated networks, we propose a preferential contact strategy based on the local network structure and local informed density to promote the information spreading. During the spreading process, an informed node will preferentially select a contact target among its neighbors, basing on their degrees or local informed densities. By extensively implementing numerical simulations in synthetic and empirical networks, we find that when only consider the local structure information, the convergence time of information spreading will be remarkably reduced if low-degree neighbors are favored as contact targets. Meanwhile, the minimum convergence time depends non-monotonically on degree-degree correlation, and moderate correlation coefficients result in most efficient information spreading. Incorporating the informed density information into contact strategy, the convergence time of information spreading can be further reduced. Finally, we show that by using local informed density is more effective as compared with the global case.
\end{abstract}

%\pacs{89.20.Hh, 89.20.Ff, 89.65.-s, 89.75.Fb}

\maketitle

%\section{Introduction} \label{sec:intro}

In the last decade, spreading dynamics in complex networks has attracted much attention from disparate disciplines, including mathematics, physics, social sciences, etc~\cite{May1992,Dorogovtsev2008,Pastor2015,Castellano2009}. Spreads of rumors~\cite{Moreno2004,Pedro2007,Damian2002}, innovations~\cite{Metcalfe1987,Strang1998,wangwei3}, credits~\cite{Banerjee2013}, behaviors~\cite{Centol2010} and epidemics~\cite{wangwei1,wangwei2} were studied both in theoretic and empirical aspects. Spreading models, such as susceptible-infected (SI)~\cite{Barthelemy2004,zhoutao2006,Vazquez2006}, susceptible-infected-susceptible (SIS)~\cite{Pastor2001prl,Pastor2001pre} and susceptible-infected-recovered (SIR)~\cite{May2001,Moreno2002,Valdez2012} have been studied to investigate the essential aspects of spreading processes in complex networks~\cite{Newman2010}. Theoretical studies revealed that underlying network structure have significant impacts on the outbreak threshold as well as outbreak size~\cite{Pastor2015}. Specially, for scale-free networks with degree exponent $\gamma\leq3$, the outbreak threshold vanishes in the thermodynamic limit~\cite{Pastor2001prl,Pastor2001pre,Boguna2013,Castellano2010}. Further studies revealed that the degree heterogeneity promotes spreading outbreaks, however limits the outbreak size at large transmission rates~\cite{wangwei1}.

Utilizing network information to effectively enhance the spreading speed and outbreak size is an important topic in spreading dynamics studies~\cite{Kitsak2010,SenPei2014,Ghoshal2011,lvLinyuan2011,Ai-XiangCui2014}. The studies on effective information spreading can provide inspiration for epidemic controlling~\cite{Funk2009,Granell2013,wangwei4}, as well as marketing strategies optimization~\cite{Bakshy2001,Watts2007,Goel2010}. Methods for effective spreading roughly fall into two categories: one is to choose influential nodes as the spreading sources~\cite{SenPei2013,liuying2015}, while the other is to employ proper contact strategies to optimize spreading paths~\cite{YangR2008}. Noticeable methods have been proposed for both the two classes. For the identification methods of influential nodes, Kitsak \emph{et~al.} revealed that selecting nodes with high \emph{k}-shells as spreading sources can effectively enhance the spreading size~\cite{Kitsak2010}. Recently, Morone \emph{et~al.} proposed an optimal percolation method to identify the influential nodes~\cite{Morone}.
As for the contact process (CP) without bias in heterogenous networks, scholars found that the spreading process follows a precise hierarchical dynamics, i.e, the hubs are firstly informed, and the information pervades the network in a progressive cascade across smaller degree classes~\cite{YangR2007}. Yang \emph{et~al.} proposed a biased contact process by using the local structure information in uncorrelated networks, and their results indicate that the spreading can be greatly enhanced if the small-degree nodes are preferentially selected~\cite{YangR2008,YangR2008_2}. Rumor spreading and random walk models with biased contact strategy were also studied in Refs.~\cite{Roshani2012, Fronczak2009}.

Previous results have manifested that in uncorrelated networks, designing a proper contact strategy can effectively promote the information spreading. However, degree-degree correlations (i.e., assortative mixing by degree) are ubiquitous in real world networks~\cite{Newman2002,Newman2003,Barabasi2016}. A positive degree-degree correlation coefficient indicates that nodes tend to connect to other nodes with similar degrees. While for negative correlation coefficients, large-degree nodes are more likely to connect to small-degree nodes. The degree-degree correlations have significant impacts on spreading dynamics. For instance, assortative (dissortative) networks have a smaller (larger) outbreak threshold, however outbreak size is on the contrast inhibited (promoted) at large transmission rates~\cite{Boguna2003,JavierB2003}.

Although correlations are prevalent in real-world systems, there still lack studies of effective spreading strategy focusing on correlated networks. To promote the information spreading in correlated networks is the motivation of this paper. We propose a preferential contact strategy based on the local information of network structure and informed densities. Our findings demonstrate that, when only consider local structure information, small-degree nodes should be preferentially contacted to promote the spreading speed, irrespective to the values of degree correlation coefficients. For highly assortative or disassortative networks, small-degree nodes should be more strongly favored to achieve the fastest spreading. Actually, the minimum convergence time of information spreading depends non-monotonically on the correlation coefficient. In addition, we find that the spreading can be further promoted when the information of informed density is incorporated into the preferential contact strategy. The local informed density based strategy can better accelerate the spreading, as compared with the global density case.

\section{RESULTS}

\textbf{Model of correlated network.} To study the interplay of degree correlations and contact strategies, we build correlated networks with adjustable correlation coefficients by employing a degree-preserving edge rewiring procedure.
First we generate uncorrelated configuration networks (UCN)~\cite{Catanzaro2005}
with power-law degree distributions and a targeted mean degree. Then, we adjust the degree correlation coefficient by using the biased degree-preserving edge rewiring procedure~\cite{Brunet2004}. Details about the network generation can be found in the Methods Section.

\textbf{Model of information spreading.} We consider a contact process (CP) of susceptible-informed (SI)~\cite{Toyoizumi2012} as the information spreading model. For the SI model, each node can either be in S (susceptible) state or I (informed) state. Initially, a small portion of nodes are chosen uniformly as informed nodes, while the remainings are in the S state. At each time step, each informed node $i$ select \emph{one} of its neighbors $j$ to contact with a pre-defined contact probability $W_{ij}$. If the node $j$ is in S state, then it will become I state with the transmission probability $\lambda$. During the spreading process, the synchronous updating rule is applied, i.e., all informed nodes will attempt to contact their neighbors in each time step~\cite{Schonfisch1999}. Repeat this process till all nodes are informed, and the network converges to an unique all-informed state. Thus for the model we consider, spreading efficiency can be evaluated by the convergence time $T$, which is defined as the number of time steps that all nodes become informed.

\textbf{Preferential contact strategy based on local information.} In real spreading processes, it is hard for nodes to known explicitly the states of neighbors. The lack of information may arise many redundant contacts between the two informed nodes in the CP, which will greatly reduce the spreading efficiency. Thus, we propose a preferential contact strategy, which combines the local structure and local informed density information in a comprehensive way. The probability $W_{ij}$ that an informed node $i$ selecting a neighbor $j$ for contact is given by
\begin{equation}\label{eq3}
W_{ij}=\frac{k_j^{\alpha+\beta\rho^L_j(t)}}{\sum\limits_{l\in \Gamma_i}k_l^{\alpha+\beta\rho^L_l(t)}}.
\end{equation}
Here $\Gamma_i$ is the set of neighbors of $i$ and $k_i$ its degree. In addition, $\alpha$ and $\beta$ are two tunable parameters. The preferential structure exponent $\alpha$ determines the tendency to contact small-degree or large-degree nodes. Large-degree neighbors are preferentially contacted when $\alpha>0$, while small-degree neighbors are favored when $\alpha<0$. When $\alpha=0$ all neighbors are randomly chosen, which reduces to the classical CP strategy~\cite{Toyoizumi2012}. The preferential dynamic exponent $\beta$ reflects whether the neighbors with small or large local informed densities are favored. For a specific node $j$,
the local informed density is defined as:
\begin{equation}\label{eq2}
\rho^{L}_{j}(t)=\frac{I_{j}(t)}{k_j},
\end{equation}
where $I_j(t)$ is the number of informed neighbors of node $j$ at time $t$.

Taking $\rho^L_j(t)$ into the contact strategy is based on several considerations. Firstly, suppose there are two neighbors with the same degree, clearly the neighbor with a higher $\rho^L_j(t)$ has a larger probability to be already informed. It is reasonable to preferentially choose the neighbor of the smaller $\rho^L_j(t)$ as contact target by setting a suitable negative $\beta$. Secondly, contacting neighbors with low informed densities can further provide more latent chances to inform the next-nearest neighbors. Third, the local informed density is relatively easier to obtain, as compared with the global informed density of network $\rho^G(t)=I(t)/N$, where $I(t)$ is the total number of informed nodes in the network at time $t$. For comparison, we also investigate the performance of global information based strategy, where the contact probability is given by replacing $\rho^L_j(t)$ with $\rho^G(t)$ in Eq.~(\ref{eq3}).

\begin{figure}
\centerline{\includegraphics[width=1\linewidth]{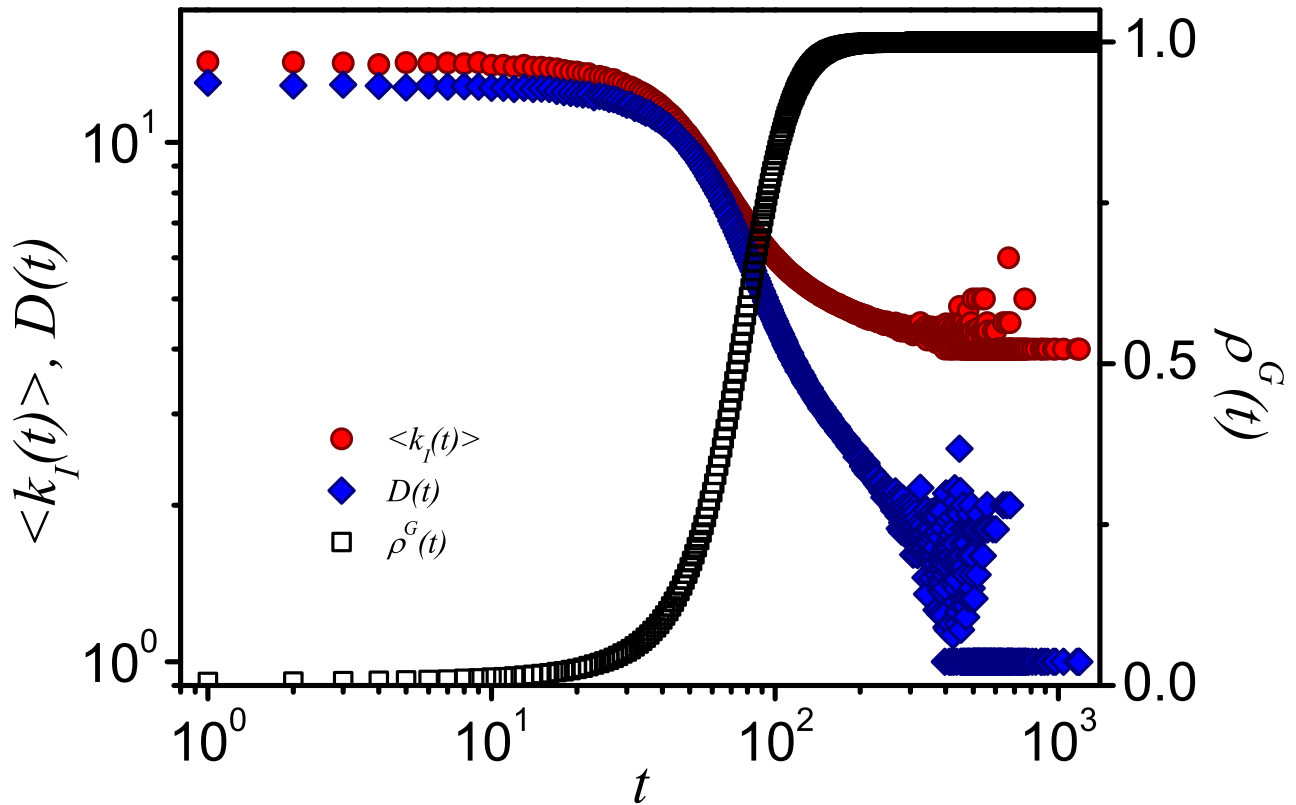}}
\caption{\textbf{For unbiased contacts, the time evolutions of information spreading in random scale-free networks.} The mean degree $\langle k_I(t) \rangle$ of newly informed nodes (red circles), the density of informed nodes $\rho^G(t)$ (black hollow squares), and the informed diversity of degrees $D(t)$ (blue diamonds) versus time steps $t$. Other parameters are set as $ N=10^4$, $\gamma=3.0$, $\langle k\rangle=8$, $\lambda=0.1$, and $\rho^G(0)=5$\textperthousand\ respectively. }\label{fig1}
\end{figure}

\textbf{Intrinsic motivation of the proposed preferential contact strategy.} We investigate the time evolutions of information spreading with unbiased contact strategy in heterogeneous random networks. When $\alpha=\beta=0$, hubs have a higher probability to be contacted since they have more neighbors. As a result, the hubs will become informed quickly. In contrast, small-degree nodes with fewer neighbors are less likely to be contacted and informed. To be concrete, the above scenario is illustrated in Fig.~\ref{fig1}. We show the time evolutions of the informed density $\rho^G(t)$, mean degree of newly informed nodes $\langle k_I(t) \rangle$, and the degree diversity of the newly informed nodes $D(t)$~\cite{Jost2007}. Here $D(t)$ is defined as:
\begin{equation}\label{eq4}
D(t)= \sum\limits_{k}\left[\frac{I_k(t)-I_k(t-1)}{I(t)-I(t-1)}\right]^{-2},
\end{equation}
where $I_k(t)$ is the number of informed nodes with degree $k$ at time $t$. The larger values of degree diversity $D(t)$ indicate that the newly informed nodes are from diverse degree classes.

At initial time steps, the informed density $\rho^G(t)$ is small, and a large value of $\langle k_I(t) \rangle$ indicates that hubs are quickly informed. The value of $D(t)$ is also very large during this stage since most nodes are in S state and nodes from all degree classes can be get informed. With the rapid increase of $\rho^G(t)$, both $\langle k_I(t) \rangle$ and $D(t)$ decreases, which indicates intermediate degree nodes are gradually informed. In the late stage of spreading, small values of $\langle k_I(t) \rangle$ and $D(t)$ reveal that the time-consuming part of the spreading is to reach some small-degree nodes. Previous studies showed that the optimal biased contact strategy basing on the neighbor degrees is which when $\alpha\approx-1$~\cite{Toyoizumi2012}. In this case, the small-degree nodes will be informed more easily, while the central role of the hubs for transmitting information is not excessively weakened. Therefore, balancing the contacts to small-degree and large-degree nodes is essential to the problem of facilitating the spreading.

Assortative and disassortative networks display distinct structure characteristics, with small-degree nodes play different roles~\cite{Newman2010}. For assortative networks, many small-degree nodes locate in the periphery of the network. While for disassortative networks, some small-degree nodes act as bridges of connecting two large-degree nodes, and with more small-degree nodes act as leaf nodes in the star-like structures.
While the locations of small degree nodes have been altered by the degree correlations, transmitting information effectively to small degree nodes is essential for facilitating the spreading as discussed above.
This suggests that we should treat small degree nodes more carefully in correlated networks. In the neighbor degree based contact strategy, nodes are identified by their degrees, and all neighbors with the same degree
are treated as equivalent. This motivates that we could further distinguish the small degree nodes to better enhance the spreading. To this end, we incorporate the local informed densities of neighbors
into the contact strategy and favor transmitting information to low informed regions.

\textbf{Simulation results for local structure information based contact strategy.} We verify the performance of the contact strategies in scale-free networks with given mean degrees and degree correlation coefficients. The networks are generated according to the method described in the model section. The size of networks is set to $N=10^4$ and average degree $\langle k\rangle=8$. In addition, we apply the method to two empirical networks£¬which are the Router~\cite{Spring2004} and CA-Hep~\cite{Leskovec2007}. Initially, $5$\textperthousand\ nodes are randomly chosen as the seeds for spreading. Without lose of generality, the transmission rate is set as $\lambda=0.1$.
All the results are obtained with averaging over $100$ different network realizations, with $100$ independent runs on each realization.

First we investigate the the interplay between degree correlations and contact strategy solely based on neighbor degrees (with $\beta=0$). Since all nodes in the network will eventually be informed for the SI model, we measure the spreading efficiency by the convergence time $T$, i.e., the time steps needed for all nodes get informed. For a specific network, there always exists an optimal value of preferential structure exponent $\alpha_o$, which will lead to the minimum convergence time $T_o$. The value of $\alpha_o$ relies on the structure of the underlying network. We focus on the relation between $\alpha_o$, $T_o$ and the degree correlation coefficient $r$.

\begin{figure}
\centerline{\includegraphics[width=0.8\linewidth]{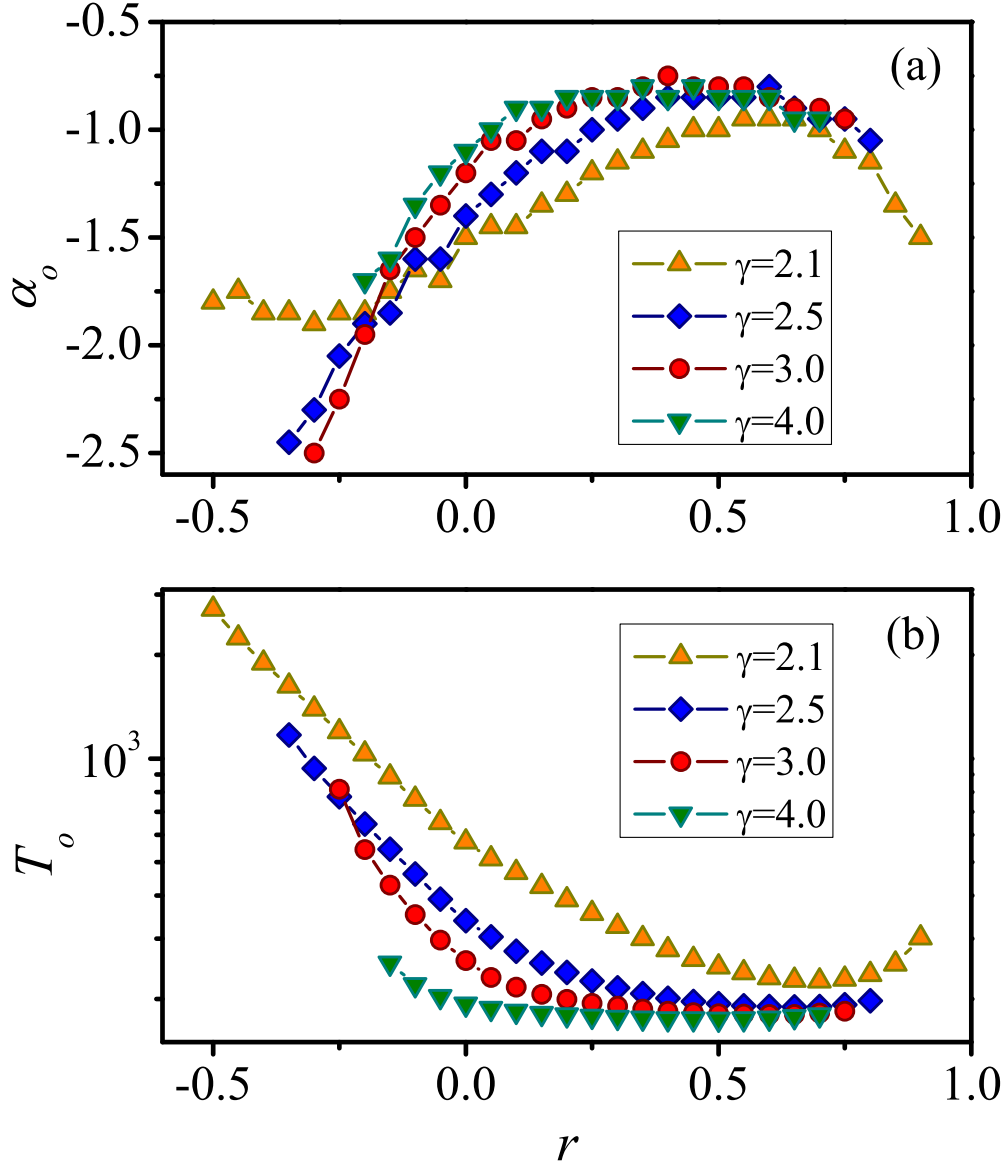}}
\caption{\textbf{The optimal performance of local structure information-based contact strategy in correlated configuration networks.} (a) The optimal value of preferential structure exponent $\alpha_o$ and (b) the convergence time $T_o$ versus correlation coefficient $r$ for degree exponents $\gamma=2.1$ (orange up triangles), $\gamma=2.5$ (blue diamonds), $\gamma=3.0$ (red circles), and $\gamma=4.0$ (green down triangles), respectively. We set other parameters as $N=10^4$, $\langle k\rangle=8$, and $\lambda=0.1$, respectively.}\label{fig2}
\end{figure}

Fig.~\ref{fig2} shows $\alpha_o$, $T_o$ versus $r$ for networks with different degree exponent $\gamma$. From Fig.~\ref{fig2}(a), we see that $r$ has significant impacts on the value of $\alpha_o$.
Generally, the values of $\alpha_o$ are negative irrespective of $r$. In other words, preferentially contacting small-degree neighbors will promote the spreading efficiency, which is consist with the previous studies for uncorrelated networks~\cite{Toyoizumi2012}. More importantly, $\alpha_o$ depends non-monotonically on $r$. In particular, when $r$ is either very large or small, $\alpha_o$ tends to be smaller than that for the intermediate values of $r$. Thus, for highly assortative and disassortative networks, it requires stronger tendency to contact small-degree nodes to achieve optimal spreading.

\begin{figure}
\centerline{\includegraphics[width=0.95\linewidth]{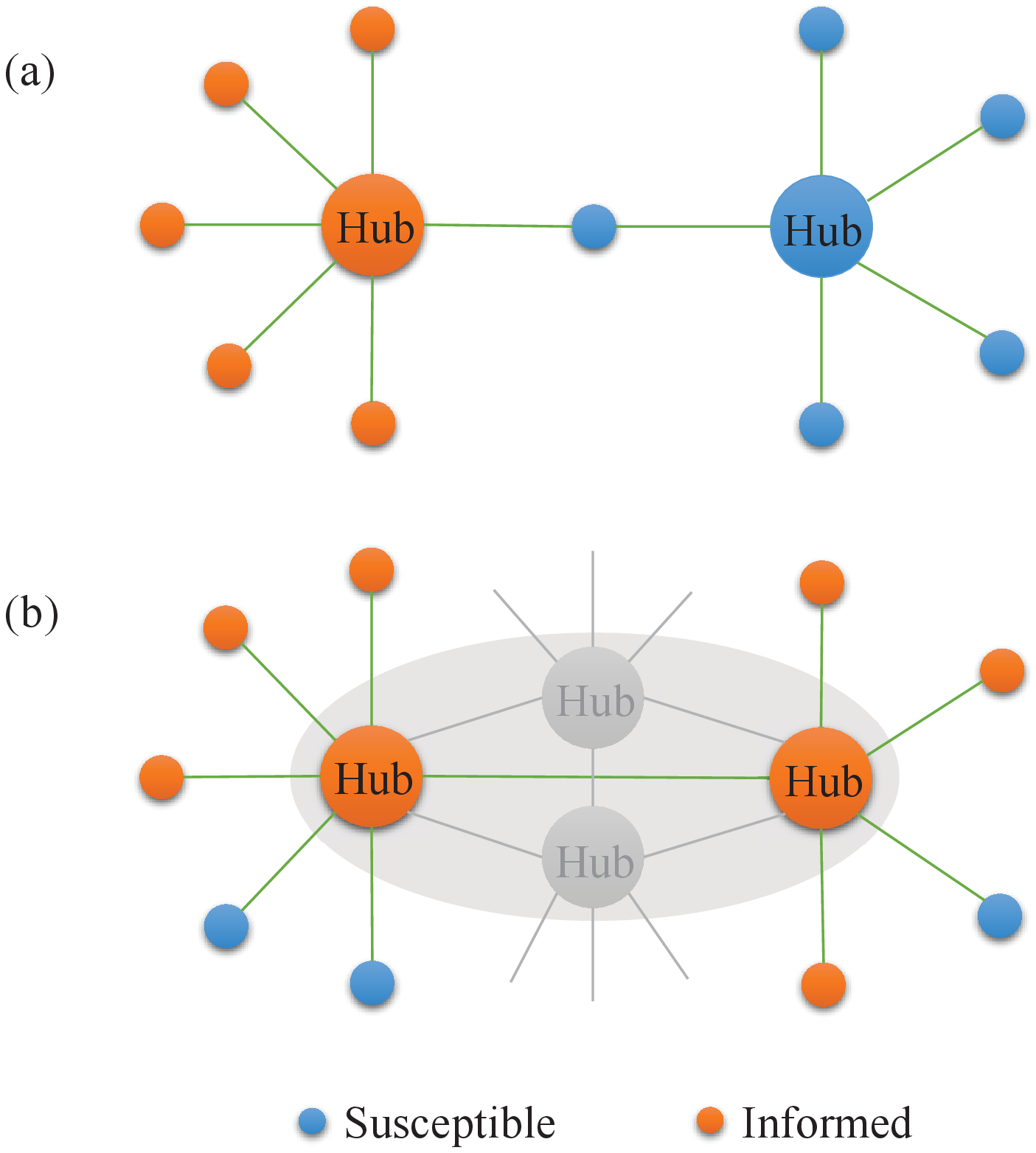}}
\caption{\textbf{Illustration of SI model in the CP for (a) disassortative networks and (b) assortative networks.} Each node can be in S (blue) state or I (orange) state.}\label{fig3}
\end{figure}

We can understand the above phenomena as follows. In extremely disassortative networks, hubs are surrounded by small-degree nodes and thus form star-like structures. Usually the hubs are not directly connect to each other and the star-like groups are interconnected via small-degree nodes. A typical such structure is illustrated in Fig.~\ref{fig3} (a). To quickly transmit the information from an informed star-like group to other groups, the small-degree nodes, which take the role of bridges, demand to be more preferentially contacted. For the extremely assortative networks, large-degree nodes form a rich club and locate in the core of networks, while small-degree nodes locate in the periphery, as shown in Fig.~\ref{fig3} (b). In this case, the information can easily spread in the core. But small-degree nodes should be preferentially contacted to avoid redundant contacts among hubs. In conclusion, both for highly assortative and disassortative networks, small-degree nodes should be more favored. When tuning the correlation coefficient of network, say from assortative to disassortative, the core-periphery structures gradually break up and turn into the star-like  structures. During this process, $\alpha_o$ first increases and then decreases. This explains the non-monotonic relationship between $r$ and $\alpha_o$.

We have also tested the method for networks with different values of degree exponent $\gamma$ in Fig.~\ref{fig2}(a). It can be seen that for different $\gamma$ the behaviors are similar. However, one noticeable difference is that for $\gamma=2.1$, $\alpha_o$ is significantly larger than the other three cases when $r$ is small. We argue that this anomaly is caused by the structural constrains imposed by the strong heterogeneity of degrees for $\gamma=2.1$. Some structural properties for $\gamma=2.1$ and $\gamma=3.0$ are summarized in table~\ref{tab1}. For $\gamma=3.0$ and $r=-0.3$, the mean degree of neighbors of the highest-degree node
$\langle k\rangle_{\mathrm{\Gamma(h)}}$ is small and close to the minimum degree of the network $k_\mathrm{min}$. In addition, we measure the degree heterogeneity of neighboring nodes of the highest-degree node
$H_{\mathrm{\Gamma(h)}}$. Low value of $H_{\mathrm{\Gamma(h)}}$ indicates that almost all the neighbors have very small degrees, which further implies the star-like structure around the hubs. On the contrary, for $\gamma=2.1$,
$\langle k\rangle_{\mathrm{\Gamma(h)}}$ is much larger than $k_\mathrm{min}$ and also the $H_{\mathrm{\Gamma(h)}}$ is of larger values. That is to say, the star-like structure around hubs is less significant for $\gamma=2.1$. In this case, small nodes do require too strong bias to achieve optimal spreading. This explains the anomaly of $\alpha_o$  for $\gamma=2.1$ and $r=-0.3$. For smaller values of $r=-0.4,r=-0.5$, $\langle k\rangle_{\mathrm{\Gamma(h)}}$ and $H_{\mathrm{\Gamma(h)}}$ also become smaller there is no very obvious star-like structure, and the $\alpha_o$ thus remains unchanged.

\label{model}

\begin{table}
  \centering
  \caption{\textbf{Some statistics of network properties for different degree exponents.} Structural properties include the mean minimum degree $k_{\mathrm{min}}$ of networks, mean degree $\langle k_{\mathrm{max}}\rangle$, mean neighboring degree $\langle k\rangle_{\mathrm{\Gamma(h)}}$ and neighboring degree heterogeneity $H_{\mathrm{\Gamma(h)}}$ of the largest-degree nodes. The neighboring degree heterogeneity is defined as
$H_{\mathrm{\Gamma(h)}}=\langle k^2\rangle_{\mathrm{\Gamma(h)}}/\langle k\rangle_{\mathrm{\Gamma(h)}}^2$, where $\langle k\rangle_{\mathrm{\Gamma(h)}}$ and $\langle k^2\rangle_{\mathrm{\Gamma(h)}}$ are the first and second moments of neighboring degrees, respectively.}
  \begin{tabular}{cccccc}
    \hline
    $\gamma$ & $r$ & $k_{\mathrm{min}}$ & $\langle k_{\mathrm{min}}\rangle$ & $\langle k\rangle_{\mathrm{\Gamma(h)}}$ & $H_{\mathrm{\Gamma(h)}}$\\
    \hline
    3& $-0.3$ & 4 & 112.5 & 4.4 & 1.031  \\%& 8
    2.1 & $-0.3$ &  2 & 134.5 & 10.3 & 4.357  \\%& 12
    2.1 & $-0.4$ &  2 & 134.5 & 5.4 & 4.024  \\%& 12
    2.1 & $-0.5$ &  2 & 134.5 & 3.1 & 2.366  \\%& 12
    %2.5 & 3 & 121.2 & 5.7 & 2.818  \\
    \hline
  \end{tabular}
  \label{tab1}
\end{table}

To further clarify the effects of correlation coefficient $r$ on the convergence time $T$, we plot the minimum convergence time $T_o$ as a function of $r$ in Fig.~\ref{fig2}(b). One can see that $T_o$ also depends non-monotonically on $r$. For those highly disassortative networks, many small clusters are interconnected via some small-degree nodes. The inter-cluster transmissions of information delay the spreading and lead to a large value of $T_o$. When $r$ is very large for assortative networks, though the core composed of large-degree nodes is easily informed, small nodes in the periphery are harder to be contacted. The core-periphery structure also gives rise to a slightly large value of $T_o$.

\begin{figure}
\centerline{\includegraphics[width=0.95\linewidth]{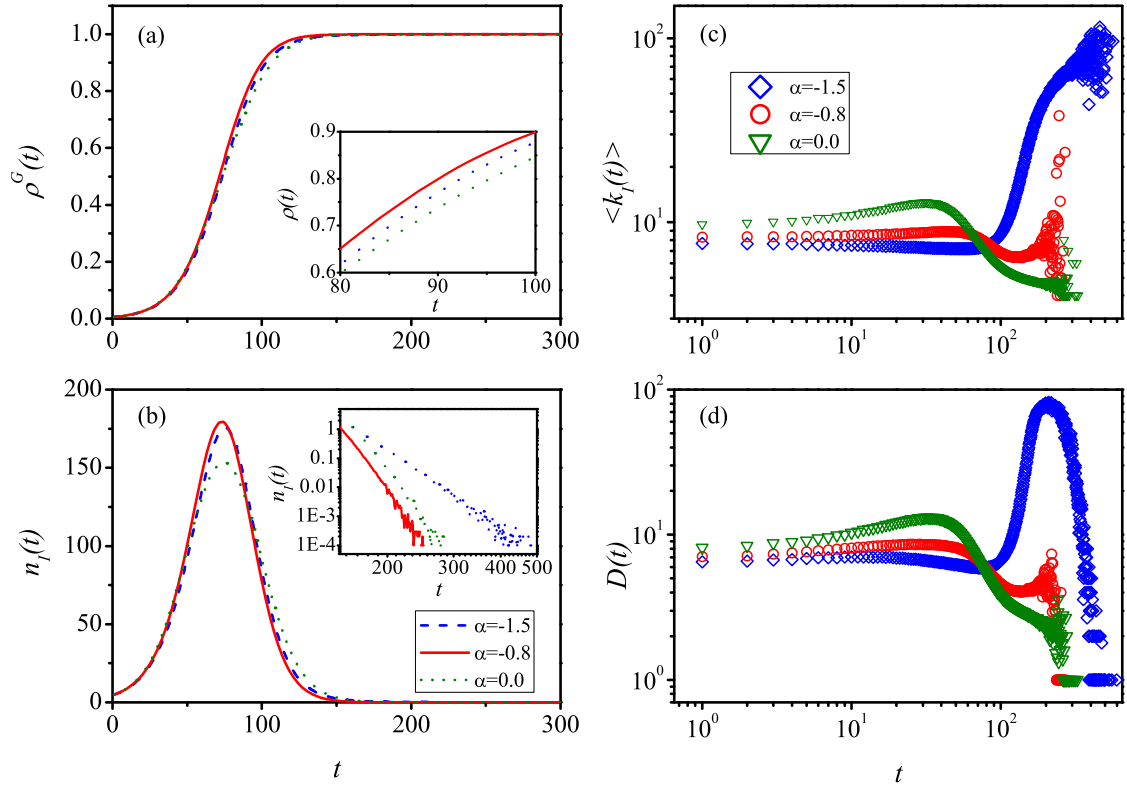}}
\caption{\textbf{The effect of local structure information-based contact strategy on the time evolution of information spreading in correlated networks.} (a) The informed density $\rho^G(t)$, and (b) the number $n_I(t)$, (c) mean degree $\langle k_I(t) \rangle$, and (d) degree diversity $D(t)$ of newly informed nodes versus $t$ for different values of $\alpha$. Different colors indicate different values of $\alpha$. The inset of (a) shows the time evolution of $\rho^G(t)$ in the time interval $[80,100]$. The inset of (b) shows $n_I(t)$ in the time interval $[150,500]$. We set other parameters as $N=10^4$, $\gamma=3.0$, $\langle k\rangle=8$, $\lambda=0.1$, and $r=0.55$, respectively.}\label{fig4}
\end{figure}

To complete the above discussions, we study the time evolution properties of the spreading process. Fig.~\ref{fig4}(a) depicts the informed density $\rho^G(t)$ versus time $t$ for the case of $r=0.55$ and $\gamma=3.0$. Note that $r=0.55$ minimize $T_o$ when $\gamma=3$, as shown in Fig.~\ref{fig2}(b). The three different lines correspond to different values of $\alpha$, which are $\alpha=-1.5,~-0.8,~0.0$, respectively. It's clear from Fig.~\ref{fig4}(a) that the case for $\alpha=-0.8$ spreads faster than the two other cases. The number of newly informed nodes $n_I(t)$ as a function of $t$ is given in Fig.~\ref{fig4}(b). One can observe that, compared with $\alpha=0$ and $\alpha=-1.5$, the $n_I(t)$ for $\alpha=-0.8$ is larger (smaller) than the other two cases at the early (late) stages, indicating the fastest spread of information. When the network is almost fully informed at late stages, the inset in Fig.~\ref{fig4}(b) demonstrates that $n_I(t)$ decays faster with time for $\alpha=-0.8$. Figs.~\ref{fig4}(c) and (d) respectively show the time evolutions of mean degree of new informed nodes $\langle k_I(t) \rangle$ and the corresponding degree diversity $D(t)$. For $\alpha=-0.8$, the $\langle k_I(t) \rangle$ and $D(t)$ remain relatively stable with time. In other words, nodes with different degrees almost have uniform probabilities of being informed, which is close to the ideal situation for effective spreading~\cite{Toyoizumi2012}. However, for $\alpha=0$ large-degree nodes are first informed and then the small-degree ones, while for $\alpha=-1.5$ the order is reversed. For the two cases, the degree diversity becomes small at the late stages of information spreading. Together with the $\langle k_I(t) \rangle$ we can conclude that the spreading is delayed by small-degree (large-degree) nodes for $\alpha=0$ ($\alpha=-1.5$). Correspondingly, the results of informed degree diversity $D(t)$ in Fig.~\ref{fig4}(d) validate the advantage of $\alpha=-0.8$ again, which is more stable than that for $\alpha=0$ and $\alpha=-1.5$.

We also apply the local structure information-based contact strategy to two empirical networks. (\romannumeral1) Router. The router level topology of the Internet, collected by the Rocketfuel Project~\cite{Spring2004}. (\romannumeral2) CA-Hep. Giant connected component of collaboration network of arxiv in high-energy physics theory~\cite{Leskovec2007}. We wish to investigate how the correlation coefficients affect the optimal value of preferential structure exponent. This is achieved by rewiring the original network with preserving the degree sequence. However, due to the abundance of degree $1$ nodes in the two empirical networks, the correlation coefficients are confined to a small region. Also, with those degree $1$ nodes it is difficult to adjust the correlation coefficients while preserving the connectivity of networks. To overcome this problem, we remove all $1$-shell nodes from the original networks~\cite{Dorogovtsev2006}. Briefly, first we remove all the nodes with degree $1$, and then re-calculate the degrees of nodes. This procedure is repeated until the degrees of all nodes are greater than one. Some structural properties of the two networks (after removing 1-shell nodes) are presented in Table~\ref{tab2}. For comparison we also list those structural properties of the randomized networks, which are obtained by unbiased degree-preserving rewiring.

Similarly, the non-monotonic dependence of $\alpha_o$ and $T_o$ on $r$ can be observed for the case of the empirical networks. Nevertheless, some abnormal bulges of $\alpha_o$ and  $T_o$ emerge at certain values of $r$. By analyzing the structures of the networks, we find that the networks are very similar to the original networks as there are few rewiring edges in the networks at these certain values of $r$ with abnormal bulges. Owing to the structural complexity of the real networks, which are significantly different from the synthetic networks, leading to abnormal bulges at certain values of $r$. To prove that the above abnormal phenomenon comes from the structural complexity of the empirical networks, we randomize the empirical networks by sufficient rewiring process but do not change the original degree distribution and the degree of each node. After sufficient times of randomization, we then check the contact strategy in the randomized networks, and one can see the abnormal bulges disappear. Moreover, the curves become more smooth and the non-monotonic phenomenon becomes more evident.

\begin{figure}
\centerline{\includegraphics[width=0.95\linewidth]{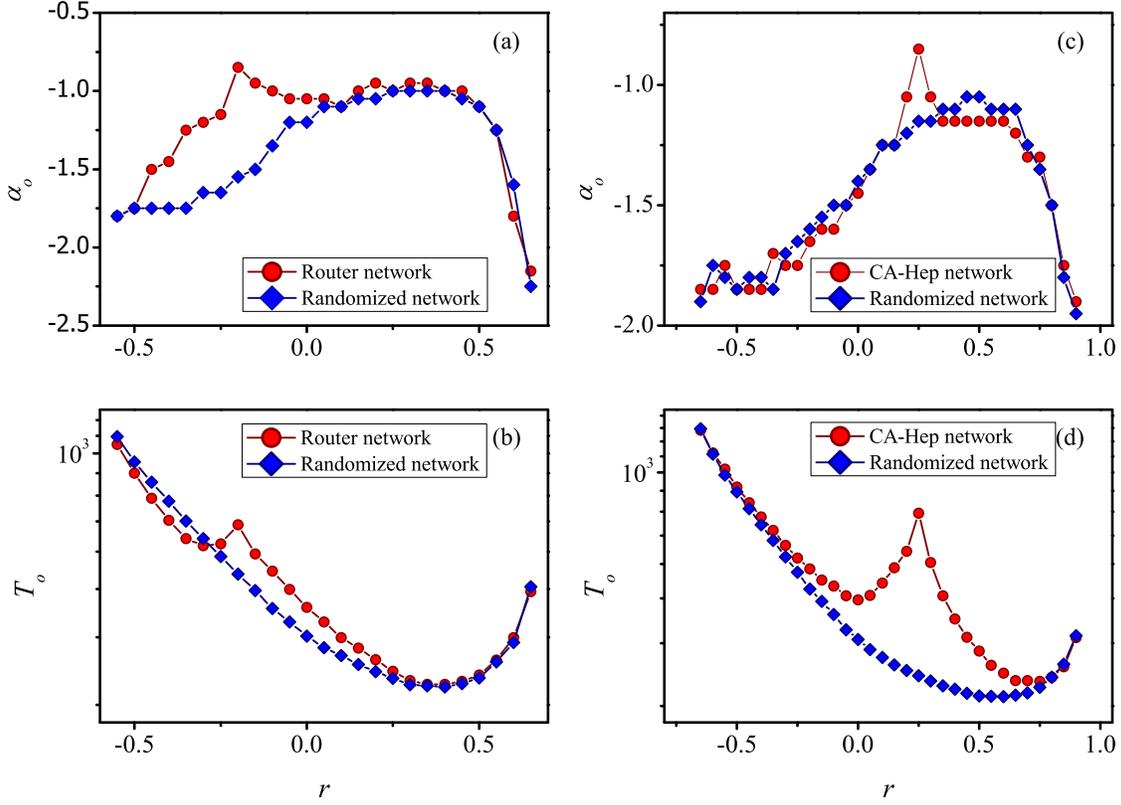}}
\caption{\textbf{The optimal performance of local structure information-based contact strategy in correlated networks by rewiring real-world networks (red circles) and randomized networks (blue diamonds).} The optimal value of preferential structure exponent $\alpha_o$ (a) and convergence time $T_o$ (b) versus correlation coefficient $r$ for the Router network. The $\alpha_o$ (c) and $T_o$ (d) versus $r$ for the CA-Hep network. }\label{fig5}
\end{figure}

\begin{table}

\caption{\label{tab2} \textbf{Structural properties of the empirical networks considered in this paper.} Structural properties include number of nodes $N$, number of edges $E$, mean degree $\langle k \rangle$, maximum degree $k_{\mathrm{max}}$, degree heterogeneity $H_k=\langle k^2 \rangle/\langle k \rangle^2$, average shortest distance $ L $, correlation coefficient $r$, and clustering coefficient $C$~\cite{Newman2010,Albert2002}.}%diameter $D$,
  \hspace{-0.9cm}
  %\centering
\begin{tabular}{c c c c c c c c c c }% c
\hline
Data &   & $N$ & $E$ & $\langle k \rangle$ & $ k_{\mathrm{max}} $& $ H_{k} $ &$  L $& $r$ & $C$ \\
\hline
\multirow{2}{*}{Router} &

  Real & 728 & 1964 & 5.4 & 59 & 2.537 & 5.232 & $-0.216$ & 0.168 \\%& 12
 & Randomized & 728 & 1964 & 5.4 & 59 & 2.537 & 3.583 & $-0.047$ & 0.040 \\%& 8
\hline
\multirow{2}{*}{CA-Hep} &

  Real & 7059 & 23227 & 6.6 & 65 & 2.001 & 5.656 & 0.247 & 0.613 \\%& 17
 & Randomized & 7059 & 23227 & 6.6 & 65 & 2.001 & 4.386 & $-0.004$ & 0.003 \\%& 9
\hline
\end{tabular}
\end{table}

\begin{figure}
\centerline{\includegraphics[width=0.95\linewidth]{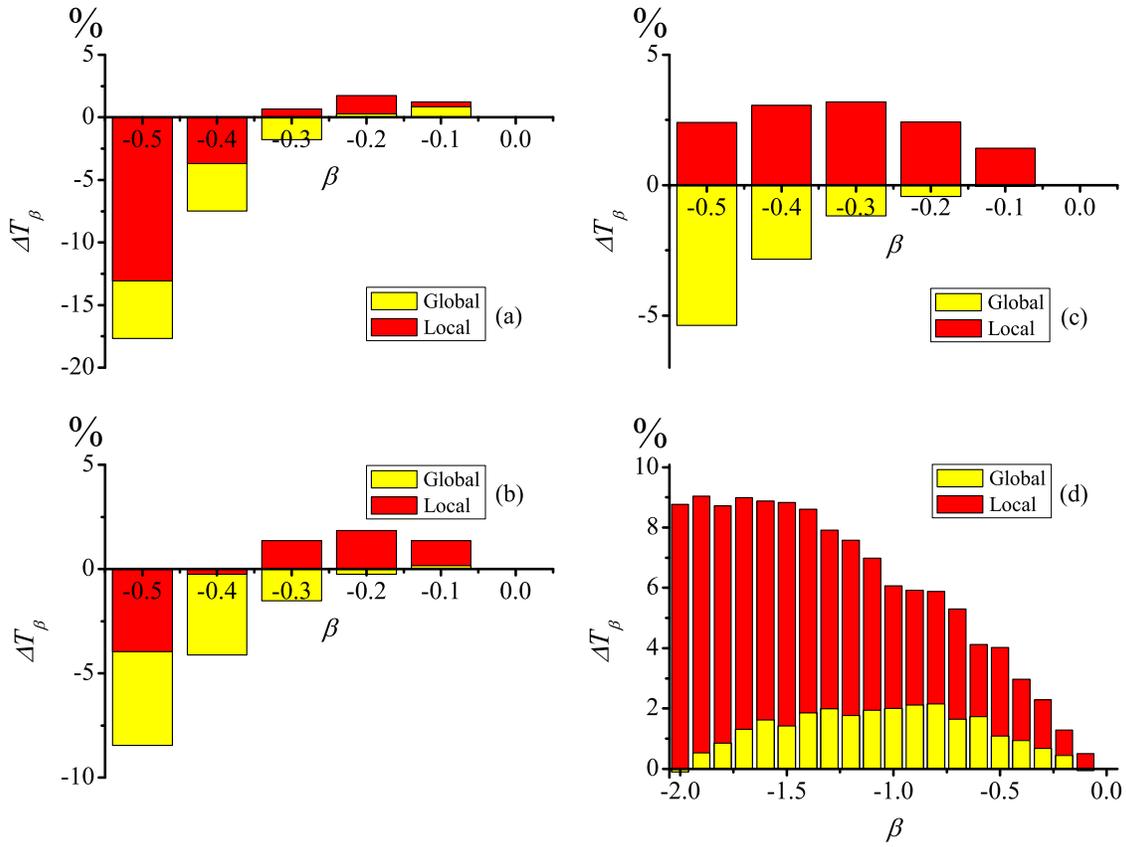}}
\caption{\textbf{In correlated networks, the effects of local and global informed density on the convergence time.} The relative ratio of the convergence time $\Delta T_\beta$ versus $\beta$ for different $r$ values: (a) $r=0$, $\alpha_o=-1.2$, (b) $r=0.55$, $\alpha_o=-0.8$, (c) $r=0.75$, $\alpha_o=-0.95$, and (d) $r=-0.3$, $\alpha_o=-2.5$, respectively. We set other parameters as $N=10^4$, $\gamma=3.0$, $\langle k\rangle=8$, and $\lambda=0.1$. }\label{fig6}
\end{figure}

\textbf{Simulation results for contact strategy with local informed density.} When $\alpha=\alpha_o$, the local structure information-based contact strategy can effectively enhance the spreading efficiency. In this section, we further incorporate local informed density information, i.e.,  with $\beta<0$ and $\alpha=\alpha_o$ in Eq.~(\ref{eq3}). The spreading efficiency $\Delta T_\beta$ is measured by $\Delta T_\beta=(T_o-T_{\beta})/T_o$, where $T_\beta$ represents the convergence time when $\beta<0$, and $T_o$ denotes the convergence time when $\beta=0$.
Thus $\Delta T_\beta>0$ ($\Delta T_\beta<0$) indicates that introducing local informed density information can enhance (inhibit) the spreading efficiency. For comparison we also study the effects of global informed density, by replacing $\rho^L_j(t)$ with $\rho^G(t)$ in Eq.~(\ref{eq3}). The results in Figs.~\ref{fig6} (a), (b) and (c) manifest that, the local informed density information can further reduce the convergence time when $\beta$ is set as a small negative value (e.g., $\beta=-0.1$ and $-0.2$). Yet, $\beta$ with larger magnitude (e.g., $\beta=-0.5$) will increase the convergence time. There is obviously an optimal value of $\beta_o$ at which the information spreading can be effectively enhanced. Moreover, compared with the global case, utilizing the local informed density information not only speeds up the spreading more significantly but also has a wider range of $\beta$ with $\Delta T_\beta>0$. For disassortative networks, as shown in Fig.~\ref{fig6}(d) with $r=-0.3$, the local informed density based contact strategy can speed up the spreading of information for a wide range of $\beta$. Such an improvement is more evident as compared to uncorrelated [Fig.~\ref{fig6}(a)] and assortative networks [Figs.~\ref{fig6}(b) and(c)]. We conclude that local informed density based strategy performs better than the global one in reducing the convergence time.

\begin{figure}
\centerline{\includegraphics[width=0.95\linewidth]{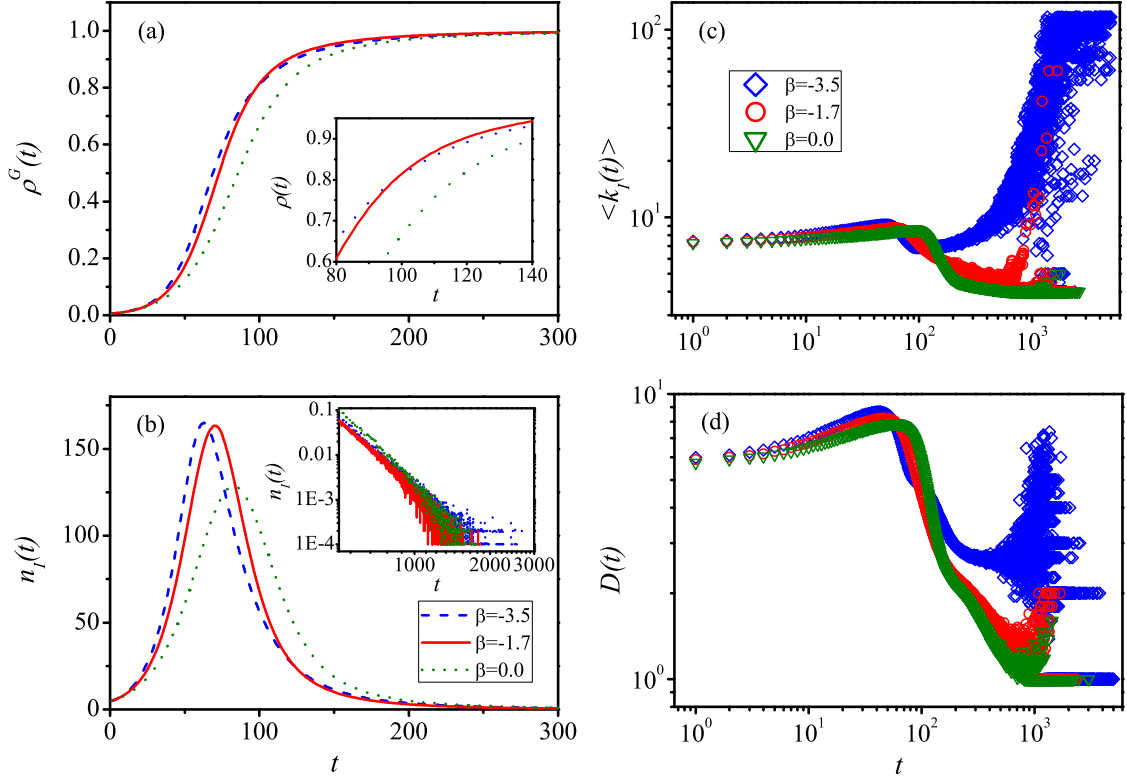}}
\caption{\textbf{The effect of local informed density on the time evolution of information spreading in disassortative networks.} (a) The informed density $\rho^G(t)$, and the node number $n_I(t)$ (b), mean degree $\langle k_I(t) \rangle$ (c) and degree diversity $D(t)$ (d) of newly informed nodes versus $t$ for different values of $\beta$. Different colors indicate different values of $\beta$. The inset of (a) shows the  time evolution of $\rho^G(t)$ in the time interval $[80,140]$. The inset of (b) shows $n_I(t)$ in the time interval $[550,3000]$. We set other parameters as $\gamma=3.0$, $N=10^4$, $\langle k\rangle=8$, $\lambda=0.1$, $r=-0.3$, and $\alpha=-2.5$, respectively.}\label{fig7}
\end{figure}

Fig.~\ref{fig7} presents time evolutions of some statistics of the spreading process in disassortative networks with $r=-0.3$. Fig.~\ref{fig7}(a) and the inset suggest that moderate $\beta=-1.7$ can better improve the speed of spreading. Fig.~\ref{fig7}(b) emphasizes that, for the case of $\beta<0$, the number of newly informed nodes $n_I(t)$ increases faster than the case of $\beta=0$ at the initial stage. However, the inset of Fig.~\ref{fig7}(b) illustrates that, for the case of $\beta=-1.7$, $n_I(t)$ goes to zero faster than the case of $\beta=-3.5$. Similar to Figs.~\ref{fig4}(c) and (d), the results in Figs.~\ref{fig7}(c) and (d) also manifest that too large (small) values of $\beta$ make the small-degree (large-degree) nodes uneasy to be informed, which will inhibit the spreading. In strongly disassortative networks, complex local structures and dynamical correlations
cause nodes with the same degree to be in different local dynamical statuses. The local structure information-based contact strategy can not effectively reflect and overcome the local dynamical status difference. Moderate values of $\beta$ guarantee the probability of being informed more homogeneous and steady for different degree classes, leading to the fastest spreading of information.

\begin{figure}
\centerline{\includegraphics[width=0.95\linewidth]{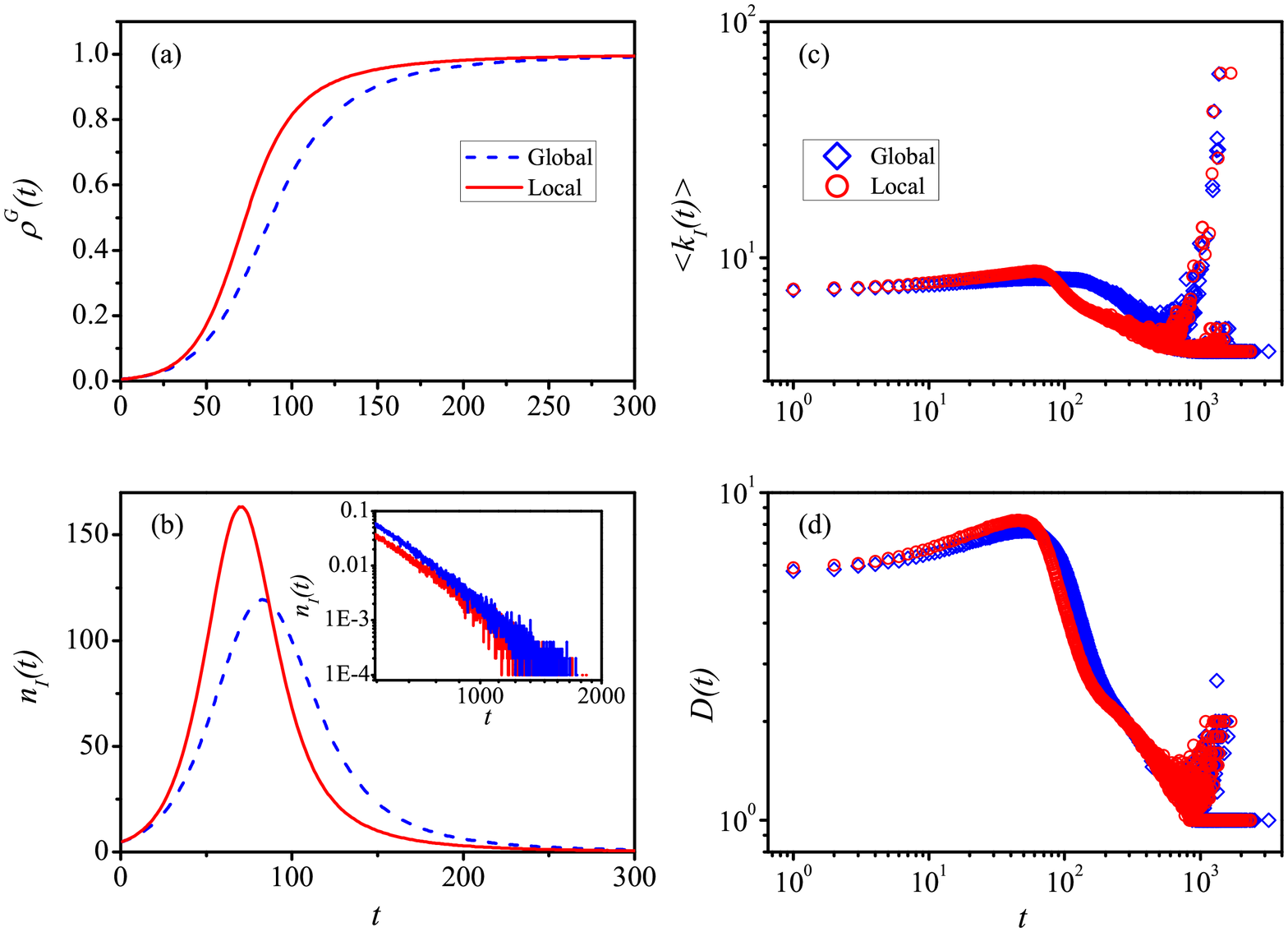}}
\caption{\textbf{Comparison of the effects of local and global informed density on the time evolution of information spreading in disassortative networks.} (a) The informed density $\rho^G(t)$, and the node number $n_I(t)$ (b), mean degree $\langle k_I(t) \rangle$ (c) and degree diversity $D(t)$ (d) of newly informed nodes versus $t$ for different values of $\beta$ by using global (color blue) or local informed density (color red). The inset of (b) shows $n_I(t)$ in the time interval $[550,2000]$. We set other parameters as
$\gamma=3.0$, $N=10^4$, $\langle k\rangle=8$, $\lambda=0.1$, $r=-0.3$, $\alpha=-2.5$, and $\beta=-1.7$ respectively. }\label{fig8}
\end{figure}

Next, we further explain why the preferential contact strategy based on the local informed density yields better performance than the global case. The time evolutions of informed density in Figs.~\ref{fig8}(a) and (b) show that the spreading speed for the local case is quicker than the global case. Since the local density information can better reveal the information distribution in a local region, some small-degree nodes with low informed density of neighbors can be informed early [see Figs.~\ref{fig8}(c) and (d)]. As a result, the information can diffused to whole networks more effectively as compared to the global case.

\begin{figure}
\centerline{\includegraphics[width=0.95\linewidth]{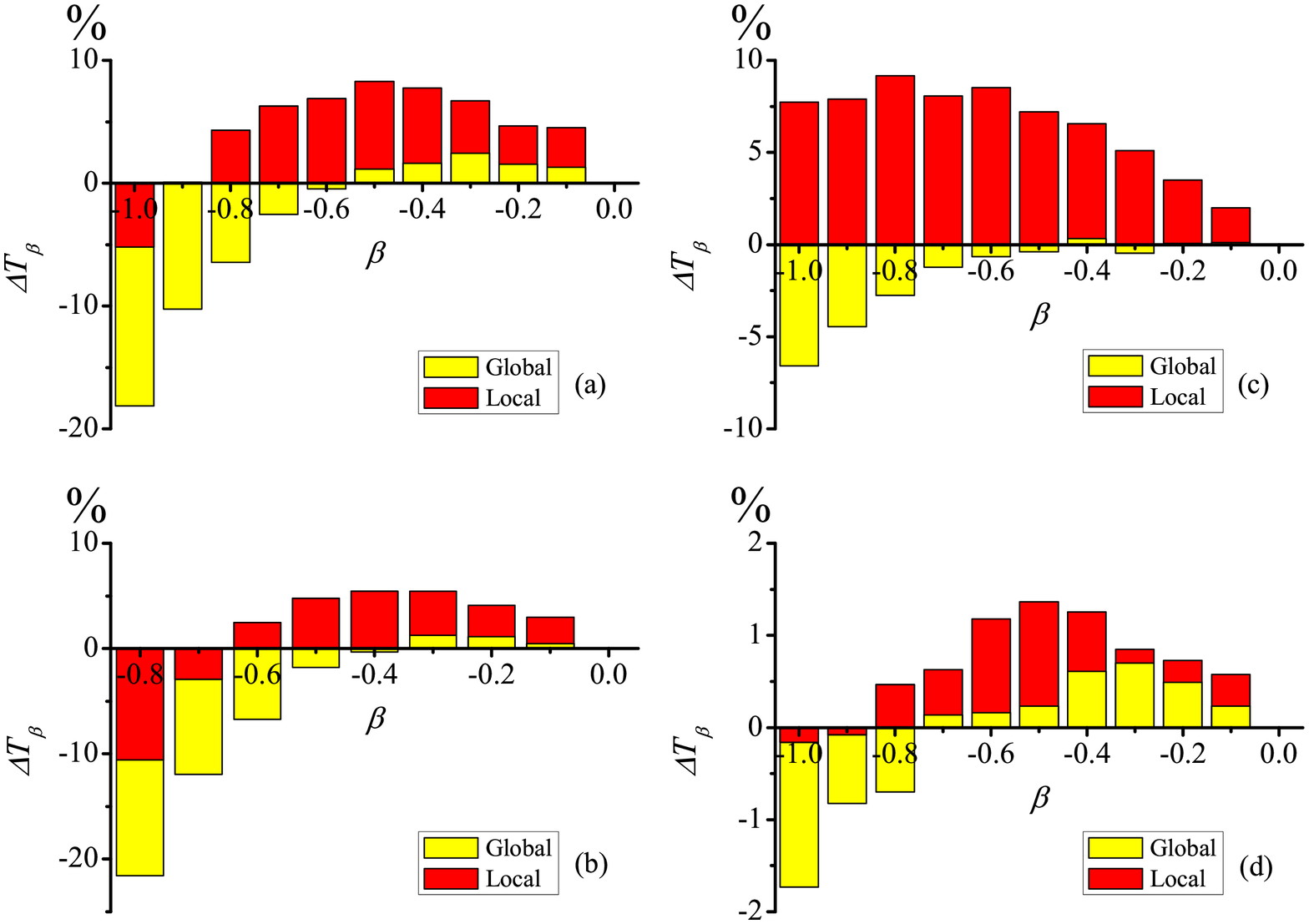}}
\caption{\textbf{Comparison of the effects of local and global informed density on the convergence time in correlated Router networks.} The relative ratio of the convergence time $\Delta T_\beta$ versus $\beta$ for different $r$ values: (a) $r=0$, $\alpha=-1.2$, (b) $r=0.4$, $\alpha=-1.0$, (c) $r=0.65$, $\alpha=-2.25$, and (d) $r=-0.55$, $\alpha=-1.8$, respectively. }\label{fig9}
\end{figure}

\begin{figure}
\centerline{\includegraphics[width=0.95\linewidth]{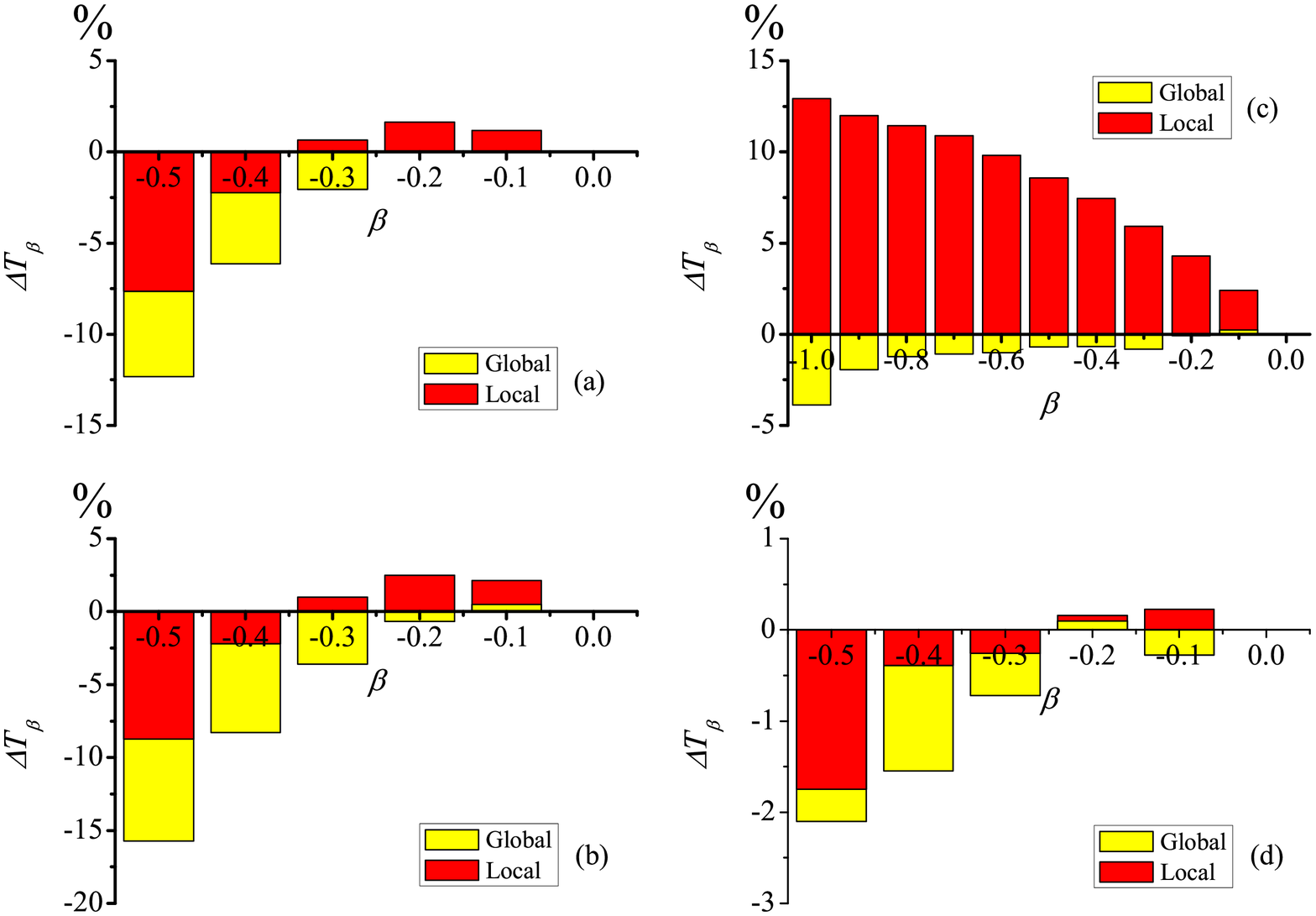}}
\caption{\textbf{Comparison of the effects of local and global informed density on the convergence time in correlated CA-Hep networks.} The relative ratio of the convergence time $\Delta T_\beta$ versus $\beta$ for different $r$ values: (a) $r=0, \alpha=-1.4$, (b) $r=0.6$, $\alpha=-1.1$, (c) $r=0.9$, $\alpha=-1.95$, and (d) $r=-0.65$, $\alpha=-1.9$, respectively.}\label{fig10}
\end{figure}

Finally, we verify the effectiveness of the informed density information based strategies in the Router network (see Fig.~\ref{fig9}) and  CA-Hep network (see Fig.~\ref{fig10}). Both of Fig.~\ref{fig9} and Fig.~\ref{fig10} show similar results with Fig.~\ref{fig6}, which is consist with above discussions. On one hand, the convergence time can be reduced when $\beta$ is slightly below zero for the local density strategy. Nevertheless, too small values of $\beta$ will instead increase the convergence time. On the other hand, introducing the local density information not only reduces the convergence time more significantly, but also yields a wider region of effective $\beta$ as compared with the global case. Thus, the local local density based contact strategy performs better in improving the speed of information diffusion.

\section{Conclusions}

To effectively promote the information spreading in correlated networks,
we proposed a preferential contact strategy by considering both the
local structure information and the local informed density. Based on
extensive simulations in artificial and real-world networks, we
verified the effectiveness of the proposed strategy, and generally
found that preferentially selecting nodes with smaller degrees and
lower local informed densities is more likely to promote information
spreading in a given network. First, we studied the strategy which
only considers the local structure information. For a given network,
there generally exists an optimal preferential exponent, at which the small-degree nodes are favored and the convergence time $T_o$ reaches its minimum value. Especially, the small-degree nodes should be favored more strongly to achieve optimal spreading when networks are highly assortative or disassortative. Also, the optimal convergence time $T_o$ depends non-monotonically to the correlation coefficient $r$.
Then, we induced the informed density into the local structure information based contact strategy with optimal exponent $\alpha_o$. Compared to the strategy with global informed density, the local density information based contact strategy reduces the convergence time more significantly.

Utilizing network information to improve the spreading is an important topic in spreading dynamics studies. In this work, we study the effect of correlated networks on the effective contact strategy basing on the local structure and informed density. Our results would stimulate further works about contact strategy in the more realistic situation of networks such as community networks~\cite{Fortunato2010,shupanpan2012}, weighted networks~\cite{Boccaletti2006,wangwei1}, temporal networks~\cite{Holme2012,Barrat2013}, and multiplex networks~\cite{Kivela2014,wangwei4}. And this work maybe provide reference for the promotion of social contagions such as technical innovations, healthy behaviors, and new products~\cite{wangwei3,Banerjee2013,Centol2010}.

\section{METHODS}
\textbf{Uncorrelated configuration model.}
We generate uncorrelated configuration networks (UCN)~\cite{Catanzaro2005}
with power-law degree distributions and targeted mean degrees as follows: (1)
A degree sequence of $N$ nodes is drawn from the power-law distribution
$P(k)\sim k^{-\gamma}$, with all the degrees confined to the region $[k_{\mathrm{min}},~\sqrt{N}]$, where $\gamma$ is the degree exponent. Note that the average of the degrees is un-controlled
but depends on $\gamma$. (2) Adjust the average of the degree sequence to a targeted value to eliminate the difference of mean degree between synthetic networks with different degree exponents~\cite{yangzimo2011}.
In detail, to transform mean degree from original mean degree $\langle k\rangle_{\mathrm{now}}$ to targeted mean degree $\langle k\rangle_{\mathrm{tar}}$, the degree of each node $i$ is re-scaled as
$k'_i=k_i\langle k\rangle_{\mathrm{tar}}/\langle k\rangle_{\mathrm{now}}$. Now the new degrees $k'_i$ may be not integers, therefore we need to convert then to integers while preserving the degree distribution and the mean degree. Since $k'_i$ can be written as $k'_i=\lfloor k'_i\rfloor+b$ with $b\in[0,1)$, we take $k'_i=\lfloor k'_i\rfloor$ with probability $1-b$, while $k'_i=\lfloor k'_i\rfloor+1$ with probability $b$. (3) The nodes with updated degrees are randomly connected via standard procedure of the UCN model.

\textbf{Adjusting degree correlation coefficient.}
We use the biased degree-preserving edge rewiring procedure to
adjust the degree correlation coefficient~\cite{Brunet2004}.
Note that this procedure is also applicable to empirical networks. The procedure is as follows: (1) At each step, two edges of the network are randomly chosen and disconnected. (2) Then we place another two edges among the four attached nodes, according to their degrees. To generate assortative (dissortaive) networks, the highest degree node is connected to the second highest (lowest) degree node, and also connect the rest pair of nodes.
If one or both of these new edges are already exist in the network, the step will be discarded and a new pair of edges will be randomly chosen. (3) Repeat this procedure till the degree correlation coefficient reaches the target value. Here the degree correlation coefficient~\cite{Newman2010} is defined as:
\begin{equation}\label{eqr}
r=\frac{\sum_{ij}(A_{ij}-k_ik_j/2m)k_ik_j}{\sum_{ij}(k_i\delta_{ij}-k_ik_j/2m)k_ik_j},
\end{equation}
where $m$ is the total number of edges in the network, $A$ is the adjacency matrix (If there is an edge between nodes i and j, $A_ij=1$; otherwise, $A_ij=0$.) and $\delta_{ij}$ is the Kronecker delta (which is 1 if $i = j$ and 0 otherwise.). When $r=0$ there is no degree-degree correlation in the network, while $r>0$ and $r<0$ indicate positive and negative degree-degree correlations respectively.

%\bibliographystyle{revtexbib}
%\bibliography{FR}

\section{Acknowledgments}
This work was supported by the National Natural Science Foundation of China (Grant Nos.~11105025, 11575041, 61473001, and 61433014), and the Fundamental Research Funds for the Central Universities (Grant No. ZYGX2015J153).

\section*{AUTHOR CONTRIBUTIONS}
L. G. and M. T. devised the research project.
L. G. and W. W. performed numerical simulations.
L. G., W. W., M. T. and L. P. analyzed the results.
L. G., W. W., M. T., L. P. and H.-F. Z. wrote the paper.

\section*{Additional information}

%{\bf Supplementary Information} accompanies this paper at

%http://www.nature.com/scientificreports

{\bf Competing financial interests}:
The authors declare no competing financial interests.

%\end{CJK*}
\end{document}